# Partial Discharge Detection on Aerial Covered Conductors Using Time-Series Decomposition and Long Short-term Memory Network

Ming Dong[1], Jessie Sun[2]

[1] ENMAX Power Corporation, Calgary, AB, Canada

[2] Auroki Analytics, Calgary, AB, Canada

Abstract

Nowadays, aerial covered conductors (CC) are increasingly used in many places of the world due to their higher operational reliability, reduced construction space and effective protection for wildlife animals. In spite of these advantages, a major challenge of using CC is that the ordinary upstream protection devices are not able to detect the phase-to-ground faults and the frequent tree/tree branch hitting conductor events on such conductors. This is because these events only lead to partial discharge (PD) activities rather than overcurrent signatures typically seen on bare conductors. To solve this problem, in recent years, ENET Centre in Czech Republic (ENET) devised a simple meter to measure the voltage signal of the electric stray field along CC, aiming to detect the above hazardous PD activities. In 2018, ENET shared a large amount of waveform data recorded by their meter on Kaggle, the world's largest data science collaboration platform, encouraging worldwide experts to develop an effective pattern recognition method for the acquired signals. For this challenge, we developed a unique method based on time-series decomposition and Long Short-Term Memory Network (LSTM) in addition to unique feature engineering process to recognize PD activities on aerial covered conductors. The proposed method is tested on the ENET public dataset and compared to various traditional classification methods. It demonstrated superior performance and great practicality.

## 1. Introduction

Today, aerial covered conductors (CC) have been increasingly used in many places around the world. Compared to bare conductors, CC has the following unique advantages [1]:

- CC can enhance system reliability by avoiding phase-to-phase fault. This is especially useful in areas with hazardous weather conditions such as strong wind where phase-to-phase short circuits often occur;
- CC can reduce the construction distances between phase conductors and between a phase conductor and the overhead carrying structures. In comparison, it is critical to maintain minimum approach distance for bare conductors for reliability and safety reasons. For CC, this requirement can be relaxed. This is especially useful in urban areas where aerial construction space can be limited;
- CC can reduce the chance of fault caused by tree or tree branch hitting the conductors. This is especially useful in areas with many trees or vegetation;
- CC can also reduce the chance of fault caused by wildlife animals such as birds or squirrels. At the same time, wildlife animals can be better protected. This is especially useful in rural areas with many uncontrolled wildlife animals.

In spite of the above significant advantages, CC has two disadvantages related to protection:

- The ordinary upstream protection devices used for bare conductor systems are mainly based on overcurrent detection and are therefore not able to detect CC's phase-to-ground fault. While this may avoid an immediate power interruption, not detecting a broken conductor fallen to the ground can create a very risky situation to people in the close proximity [2-3]. For utility companies, it is desired to detect this kind of event and take proper field measures for safety purpose.
- The tree or tree branch hitting the conductors will not be detected by the ordinary upstream protection devices. This can create a long-term potential reliability threat in some locations where the tree and tree branches frequently hit the conductors or continuously push and bend the conductors. Eventually, the power line will be damaged and broken, causing a much longer power outage or even starting a tree fire [2-3].

To solve the above problems, researchers turned to study other types of signatures rather than the traditional overcurrent signature. It was found that the above two types of activities can cause partial discharge (PD) activities [2-3]. Therefore, if the corresponding PD activities can be effectively detected, the above protection problems can be resolved.

In order to acquire PD signals on CC, researchers also looked into designing proper measurement devices. For example, [3] designed a measurement device using Rogowski coil; ENET Centre in Czech Republic (ENET) designed a simple measurement device that can be attached to the jacket of a CC to measure PDs [4-5]: a single layer coil is wrapped around the CC to acquire the voltage signal of electric stray field along the CC. In comparison with [3], ENET's meter is more cost effective. However, to make it really work, an effective pattern recognition method is required to recognize PD activities from its acquired signals. For this purpose, in 2018, ENET published a dataset which contains a large number of waveform measurements recorded using their meter. The dataset was published on Kaggle, the world's largest data science collaboration platform [5]. The intent is to attract worldwide researchers to study the waveform data and develop effective pattern recognition algorithms for PD detection. This paper elaborates a unique method that was developed and tested on this public dataset using Time-Series Decomposition and Long Short-term Memory Network.

The flowchart of the proposed method is shown in Figure 1: first, Seasonal and Trend decomposition using Loess (STL) is employed as a time-series decomposition technique to decompose every raw voltage signal into three components, i.e. trend, seasonal and residual components. The proposed method then focuses on the residual components of each signal where irregular noise such as PD is most abundant. In order to cover a wide range of information and enhance robustness, four different STL modules with four different seasonal-window lengths are applied to produce four different residual components. At the step of feature engineering, each whole-cycle residual component is divided to different time steps; for each time step, three features are engineered to characterize the signals; a special noise reduction method is used to eliminate noises that are less likely caused by PD; then for every time step, twelve features from four residual components are merged and normalized together to generate a sequential feature vector; oversampling technique is also adopted to balance out the records with PD activities and the records with no PD activities. In the end, many-to-one sequential records are constructed and are fed into a long short-term memory network (LSTM) classifier. Finally, the LSTM classifier is trained and can be used to recognize new PD signals acquired by the meter effectively.

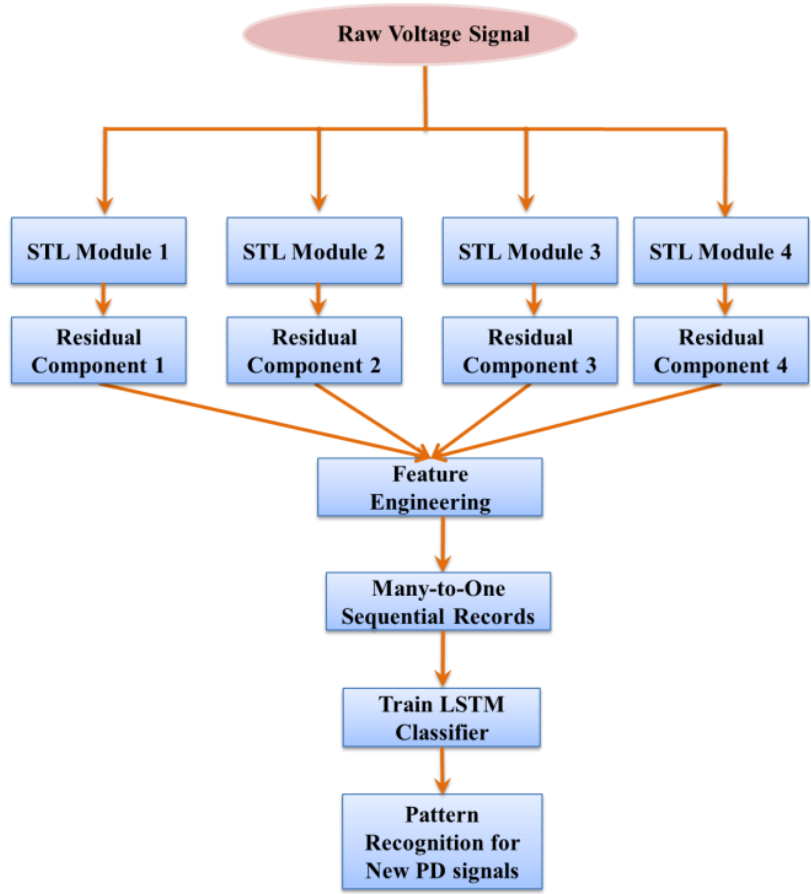

Figure 1. The flowchart of the proposed method

This paper firstly introduces the ENET dataset. It then explains the theories of time-series decomposition and LSTM network. It further elaborates the proposed step of feature engineering. In the end, the proposed method is applied to the ENET dataset. Detailed classification results and case studies are presented. The proposed method is compared with various traditional classification methods and demonstrated superior performance.

## 2. The ENET Dataset

The ENET dataset contains 8,711 pre-labeled voltage signals recorded by their meter introduced in Section 1 [5]. Each signal has been pre-labeled as either having PD (525) or no PD (8,186). Each signal is a 50Hz one-cycle voltage waveform and contains 800,000 data points. Due to the large size of this dataset, Hadoop Distributed File System (HDFS) storage format is used. Special data processing method is required to access all the signals [6]. Examples of a PD signal and a non-PD signal are plotted in Figure 2:

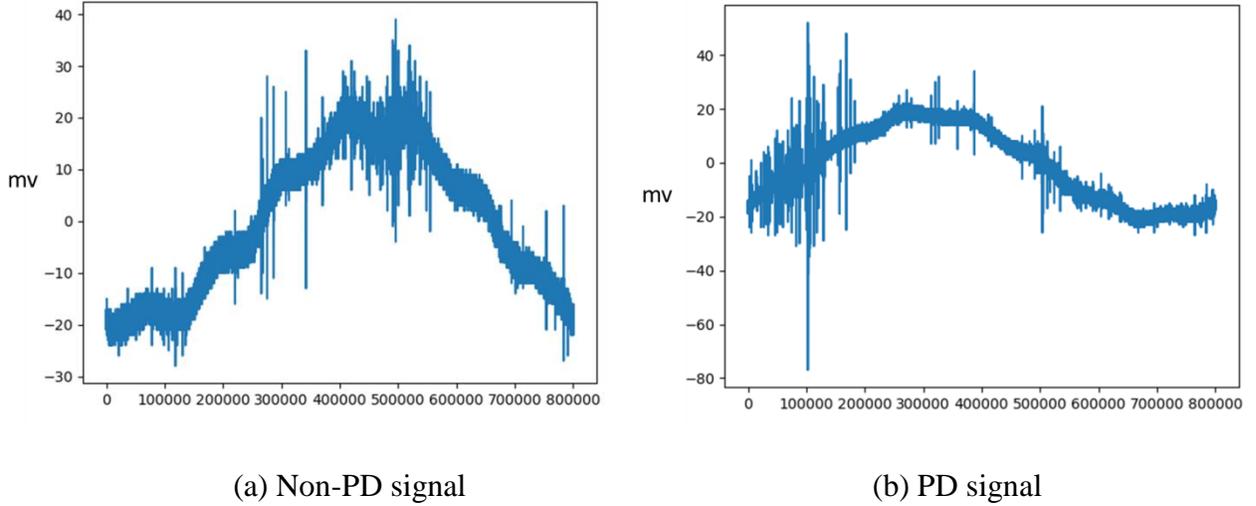

(a) Non-PD signal  (b) PD signal

Figure 2. Examples of signal records from ENET dataset

## 3. Time-Series Decomposition

The purpose of time-series decomposition is to reveal a few underlying patterns so that each pattern displays a certain characteristic or type of behavior of the time-series data. As can be seen from Figure 2, the raw signal contains noises imposed on the sinusoidal cycle. Since PD by its nature is a type of irregular noise caused by unpredictable external events such as tree branch hitting the conductor, this paper intends to firstly use time-series decomposition to extract noise information from the signals and continue to develop our method based on the extracted noise information.

STL is a widely-used time series decomposition technique [7]. It intends to decompose a time-series into three components and is mathematically given as below:

$$y_t = T_t + S_t + I_t \qquad (1)$$

where $y_t$ is the target time series for decomposition; $T_t$, $S_t$ and $I_t$ are the trend component, seasonal component and residual component respectively. Their meanings are explained as below:

- $T_t$ reflects the long-term progression in the raw signal. A trend exists when there is a persistent increasing or decreasing direction in the data. For the discussed one-cycle signal, $T_t$ can represent the slow-varying 50Hz sinusodial cycle.

- $S_t$ reflects seasonality (seasonal variation) in the raw signal. A seasonal pattern is the repeated cyclic pattern embedded in the raw signal. In the discussed signal, $S_t$ can represent the aggregated harmonic content, radio inteference and other content repeating periodically.

- $I_t$ reflects irrgularity or noise in the raw signal. It is related to random and iregualr influences. It is the residual part after excluding $T_t$ and $S_t$ from $y_t$. By nature, PD is a random and irregular component and this is the reason the proposed method focuses on the characteristics of residual component only.

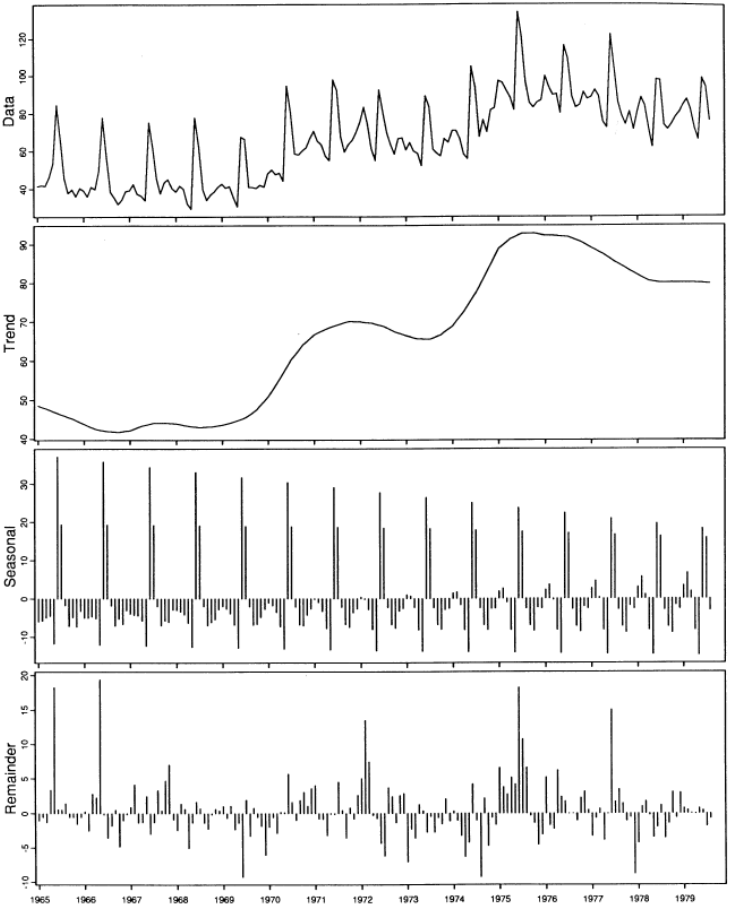

Figure 3. An example of decomposing U.S unemployed males at ages 16-19 (Y-axis unit is 10,000)

An example of using STL to decompose U.S unemployment data is shown in Figure 3 [7]. From the algorithm perspective, STL is done using Loess smoothing. Loess smoothing is a convolutional smoothing filter taking different time points around time point $t$ into consideration by applying different weighting factors. Time points far away from $t$ will be assigned with smaller weighting factors and vice versa. An inner loop, outer loop and a low-pass filter are used in conjunction with Loess smoothing to complete the decomposition process as discussed in [7].

## 4. Long Short-term Memory Network

The proposed method uses LSTM network as the classifier. Different from many traditional machine learning models such as feed-forward neural network (FNN), LSTM is a sequence prediction model. Sequence prediction is the problem of using information embedded in a series of times steps to predict a certain output [8]. In recent years, researchers applied LSTM to classic time-series problems such as stock, weather forecasting and machine translation [9-11]. They often outperform traditional machine learning models such as FNN in these tasks. The studied signal in Figure 2 can be divided into a few consecutive time steps such as half a cycle, quarter cycle and 1/8 cycle. Then LSTM can be applied to capture the sequential characteristics between these time steps. This is potentially powerful for PD detection because a PD activity may span over a few time steps with in a cycle and therefore display certain type of feature variations from one time step to the next. LSTM is able to capture and process such variations for the purpose of pattern recognition. This is the fundamental rationale of using LSTM as the classifier in the proposed method. The effect of LSTM is further tested and presented in Section 6.

This section explains the mathematical foundation of the LSTM neural network. Since LSTM neural network is essentially an enhanced recurrent neural network (RNN), this section firstly reviews standard RNN and then explains the working principle and advantages of LSTM compared to the standard RNN.

*4.1. Recurrent Neural Network*

A RNN can be viewed as a group of FNNs where hidden neurons of the FNN at the previous time step are connected with the hidden neurons of FNN at the following time step. The state of hidden neurons $H_t$ is generated from $H_{t-1}$ at the previous time step and the current input $X_t$ by applying weights $W_h$ and $W_x$. This process continues for the next time step and so on. This way, RNN is able to make use of sequential information and does not consider one time step as an isolated point. This nature made RNN suitable for forecasting tasks such as stock, weather and load forecast where the output of current time step is not only based on the current input but also the information from previous time steps. An unfolded many-to-one RNN structure is shown in Figure 4.

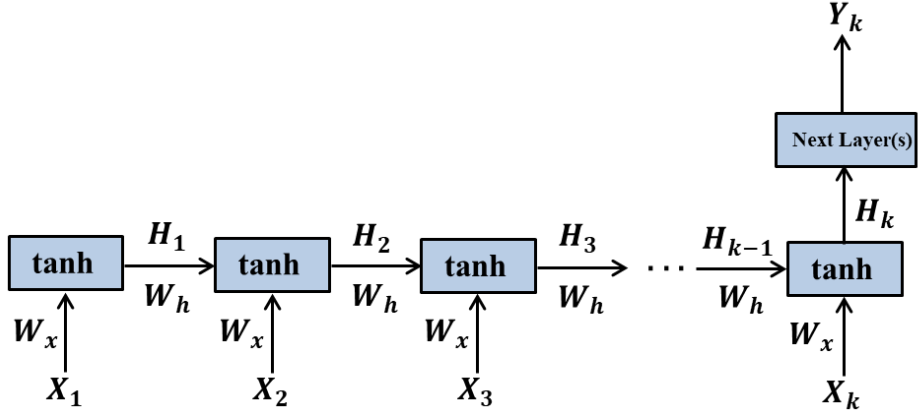

Figure 4. Illustration of an unfolded many-to-one RNN

$X$ is the input; $H$ is the hidden state vector:

$$H_t = tanh(W_h H_{t-1} + W_x X_t) \tag{2}$$

According to the chain rule, the network loss gradient should be:

$$\frac{\partial E_k}{\partial W} = \frac{\partial E_k}{\partial H_k} \frac{\partial H_k}{\partial H_{k-1}} \cdots \frac{\partial H_2}{\partial H_1} \frac{\partial H_1}{\partial W} = \frac{\partial E_k}{\partial H_k} \left( \prod_{t=2}^{k} \frac{\partial H_t}{\partial H_{t-1}} \right) \frac{\partial H_1}{\partial W} \tag{3}$$

The derivative

$$\frac{\partial H_t}{\partial H_{t-1}} = tanh'(W_h H_{t-1} + W_x X_t) \cdot \frac{d}{dH_{t-1}} [W_h H_{t-1} + W_x X_t] = tanh'(W_h H_{t-1} + W_x X_t) \cdot W_h \tag{4}$$

Combining (3) and (4), we get:

$$\frac{\partial E_k}{\partial W} = \frac{\partial E_k}{\partial H_k} \left( \prod_{t=2}^{k} tanh'(W_h H_{t-1} + W_x X_t) \cdot W_h \right) \frac{\partial H_1}{\partial W} \tag{5}$$

Apparently, $tanh < 1$. Depending on the value of $W_h$, $tanh'(W_h H_{t-1} + W_x X_t) \cdot W_h$ may be smaller than 1 which will cause gradient reduction or bigger than 1 which will cause gradient increase. Even if $K$ is not large (such as a few steps in our case), the final gradient may quickly get distorted due to the multiplication effect, depending on the random weight initialization. This is the reason RNN can be unstable and susceptible to initialization. Also, a much smaller or larger gradient can significantly affect the weight update of the network, making it hard to converge. This problem is called vanishing and exploding gradient and has been discussed in greater details in [12-13]. LSTM on the other hand replaces the tanh activation with sophisticated gate control in the gradient flow and hence can achieve better stability and performance.

*4.2. LSTM Neural Network*

As discussed above, LSTM network was proposed to improve the RNN structure [14-15]. Compared to traditional RNN, LSTM introduces a specially designed LSTM unit to sophisticatedly control the flow of hidden state information from one time step to the next. The structure of a LSTM unit is shown in Figure 5.

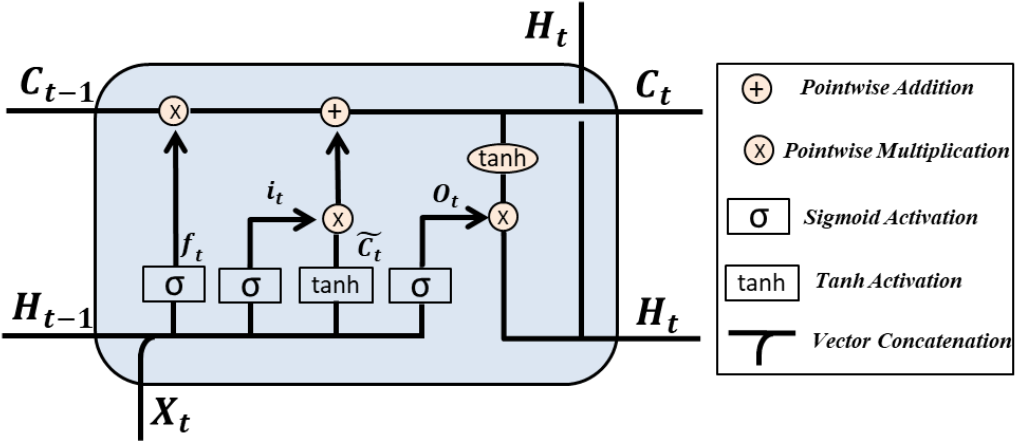

Figure 5. A LSTM Unit Diagram

In Figure 5, $X_t$ and $H_t$ are the input vector and network hidden state vector at time step $t$. $C_t$ is a vector stored in an external memory cell. This memory cell carries information between time steps, interacts with $X_t$ and $H_t$ and gets updated from one time step to the next. The interaction between cell state vector, input vector and hidden state vector is completed through three control gates: forget gate, input gate and output gate.

The forget gate vector $f_t$ is calculated by:

$$f_t = \sigma(W_f \cdot [H_{t-1}, X_t] + b_f) \tag{6}$$

where $[H_{t-1}, X_t]$ is the concatenated vector of previous hidden state vector $H_{t-1}$ and the current input vector $X_t$; $W_f$ and $b_f$ are the weights and biases for $f_t$ and are determined through network training; $\sigma$ is the sigmoid activation function. This calculation outputs a vector $f_t$. Each element in $f_t$ controls how the information in cell state vector $C_t$ can be kept. This is achieved by pointwise multiplying $f_t$ by $C_t$ and is mathematically given later in (9).

Following the information flow in Figure 5, a temporary cell state vector $\widetilde{C}_t$ is calculated by:

$$\widetilde{C}_t = tanh(W_c \cdot [H_{t-1}, X_t] + b_c) \qquad (7)$$

where $[H_{t-1}, X_t]$ is the concatenated vector of previous hidden state vector $H_{t-1}$ and the current input vector $X_t$; $W_c$ and $b_c$ are the weights and biases for $\widetilde{C}_t$; $tanh$ is the tanh activation function.

In parallel with calculating $\widetilde{C}_t$, the input gate vector $i_t$ is calculated by:

$$i_t = \sigma(W_i \cdot [H_{t-1}, X_t] + b_i) \qquad (8)$$

where $W_i$ and $b_i$ are the weights and biases for $i_t$ and are determined through network training. This calculation outputs a vector $i_t$.

Eventually the new cell state $C_t$ at time step $t$ is updated by both forget gate and input gate using pointwise multiplication:

$$C_t = f_t * C_{t-1} + i_t * \widetilde{C}_t \qquad (9)$$

This new cell state further determines the hidden state in the current neural network at time step $t$ through the write gate $o_t$. Similar to $f_t$ and $i_t$, $o_t$ is calculated by:

$$o_t = \sigma(W_o[H_{t-1}, X_t] + b_o) \qquad (10)$$

Then, hidden state $H_t$ at the current time step $t$ is calculated by pointwise multiplying $o_t$ by $tanh(C_t)$:

$$H_t = o_t * tanh(C_t) \qquad (11)$$

Through (6) to (11), the current hidden state $H_t$ is calculated with the use of $C_{t-1}$ and $H_{t-1}$ from the previous time step as well as the current input $X_t$. $H_t$ is then used by the neural network to calculate output at the current time step.

LSTM neural network inherits the advantages of RNN in dealing with temporal forecast problems and also solves the vanishing/exploding gradient problem. It is therefore chosen as the ideal mathematical model to analyze PD feature variations between different time steps within a cycle.

*4.3. Sequential Model*

As a RNN neural network, LSTM has three different sequential configurations: one-to-many, many-to-many and many-to-one [9-10]. For the studied problem, we intent to use multiple time steps to determine a binary output, i.e. whether the signal contains PD or not. Therefore, the many-to-one sequential configuration is selected for modelling. Its schematic is shown in Figure 6.

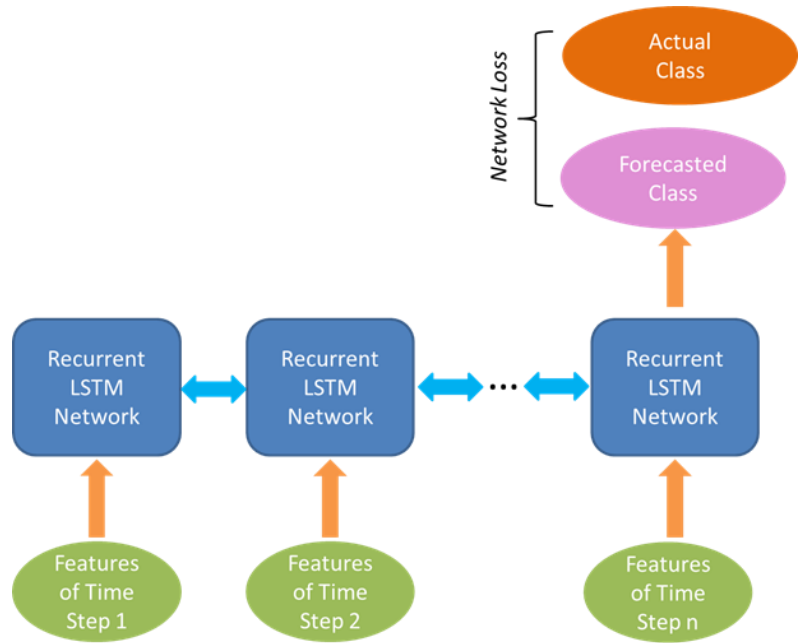

Figure 6. Many-to-One sequential model

## 5. Feature Engineering

At the step of feature engineering, each whole-cycle residual component is divided to different time steps; for each time step, three features are engineered to capture the signal characteristics: the number of peaks, the sum of absolute peak heights and the standard deviation of absolute peak heights. A special noise reduction method is used to eliminate noises that are less likely caused by PD; then for every time step, STL with four different lengths of seasonal windows is applied to each waveform signal to produce four different STL decompositions. The lengths of the seasonal windows are set to be 100, 1000, 10000 and 100000 points. Twelve features from four residual components are merged and normalized together to produce a sequential feature vector; oversampling technique is used to balance out the records with PD activities and the records with no PD activities.

This section firstly introduces the method of noise reduction. It then elaborates the steps of feature construction and fusion. In the end, the technique of oversampling is explained in detail.

*5.1. Noise Reduction*

Noise reduction is used to eliminate noises that are less likely caused by PD and extract more relevant peak signals. Two types of noises are considered in this method and are shown in Figure 7: one is the excessively large peaks which are probably transient signals caused by switching operations and lighting strikes. They are filtered by removing the largest 5% data points.

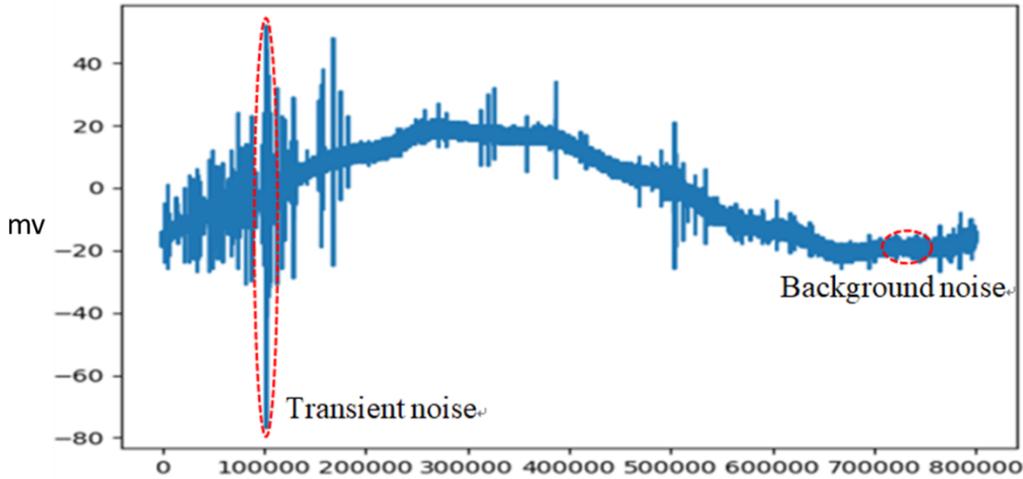

Figure 7. An example of transient noise and background noise

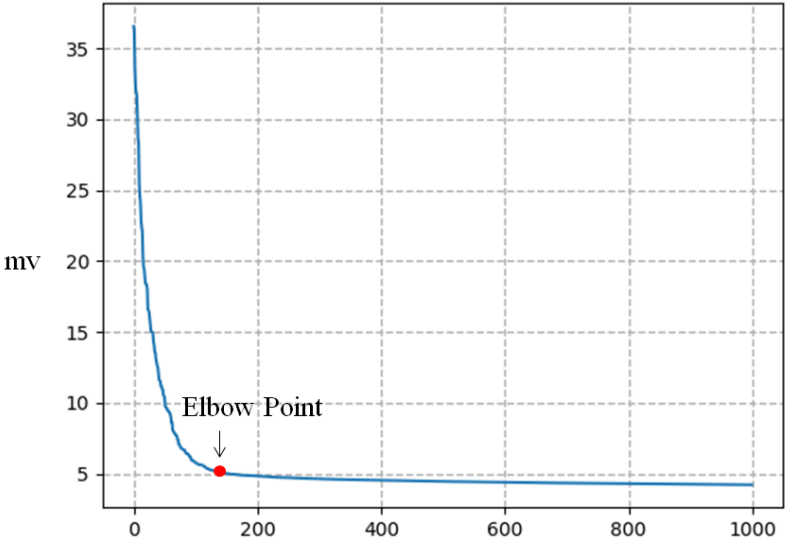

Figure 8. Elbow detection for background noise elimination (showing the first 1000 points only)

The background noise can be effectively filtered through elbow point detection. This is carried out by sorting the absolute values of data points in a residual component in descending order and searching for the area (elbow) where the values start to flatten off. Once the elbow point is located, all the points on the right side of it can be zeroed out. An example of elbow point is shown in Figure 8.

*5.2. Feature Construction*

To characterize the residual component, three features are proposed: the number of peaks, the sum of absolute peak heights and the standard deviation of absolute peak heights. Absolute values are used because there is no actual pattern difference between the positive and negative values in the residual component.

The sum of absolute peak heights $S$ is given by (12):

$$S = \sum_{i=1}^{n}|r_i| \tag{12}$$

The standard deviation of absolute peak heights $SD$ is given by (13):

$$SD = \sqrt{\frac{1}{n-1}\sum_{i=1}^{n}(|r_i| - \overline{|r|})^2} \tag{13}$$

where $n$ is the number of peaks; $r$ is the value of each peak; $\overline{|r|}$ is the average absolute value of all peaks.

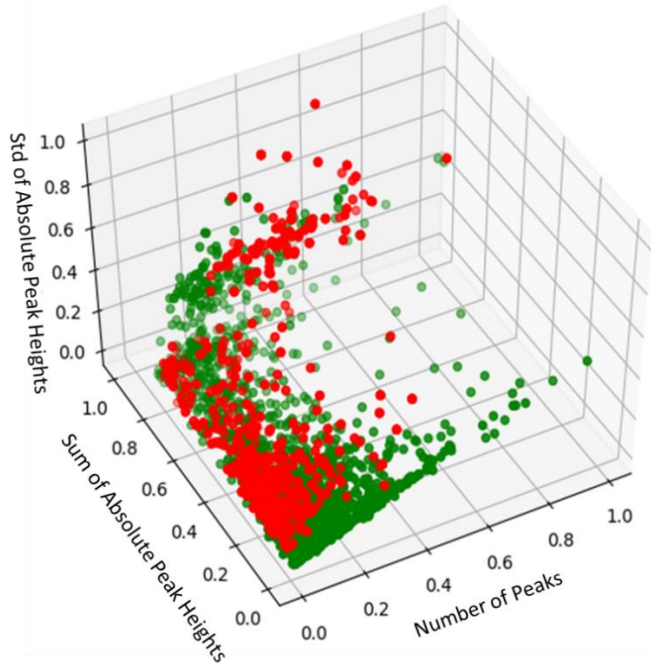

Figure 9. 3D-Plot of PD signals (red) and non-PD signals (green)

Figure 9 shows a 3D plot of the three features of 1000 randomly selected signals in the first quarter-cycle. As can been seen, intuitively many of the PD signals (red) are above the non-PD signals (green).

*5.3. Feature Fusion*

As mentioned in the beginning of this section, for each time step, three features are engineered to capture the noise characteristics: the number of peaks $n$, the sum of absolute peak heights $S$ and the standard deviation of absolute peak heights $SD$. STL with 4 seasonal-window lengths is applied to each waveform signal to produce 4 different sets of STL decomposition. All twelve features from 4 STL modules are connected as a 12-dimension sequential feature vector $f_v$ and normalized to the range of [0,1] by Max-Min scaling to avoid magnitude biases [16]. It is represented as below:

$$f_v = (n_{100}, n_{1000}, n_{10000}, n_{100000}, S_{100}, S_{1000}, S_{10000}, S_{100000}, SD_{100}, SD_{1000}, SD_{10000}, SD_{100000}) \quad (14)$$

*5.4. Oversampling for Unbalanced Dataset*

One necessary technique that is required during the training of LSTM classifier is oversampling of PD records. This is because the numbers of PD signals and non-PD signals in the dataset are very unbalanced (PD signals are much fewer than non-PD signals). If directly using the records for training, the trained classifier could result in biased classification towards non-PD signals. One technique that can overcome this problem is purposely duplicating the PD signals so that the numbers of PD signals and non-PD signals are approximately the same in the dataset. This step can ensure a classifier that can only effectively detect the majority status (non-PD signals) yields a large error. With balanced dataset, the classifier will become less biased through the training process. As discussed in [17], this technique can effectively improve the training accuracy of binary classifiers.

## 6. Result and Case Studies

*6.1. Results of Time-series Decomposition*

The proposed method shown in Figure 1 was applied to all 8,711 signals included in the ENET dataset. STL with four different lengths of seasonal windows is applied to each waveform signal to produce four different STL decompositions. The lengths of the window are set to be 100, 1000, 10000 and 100000 points. An example of the decomposition result with 1000-point window length is shown in Figure 10.

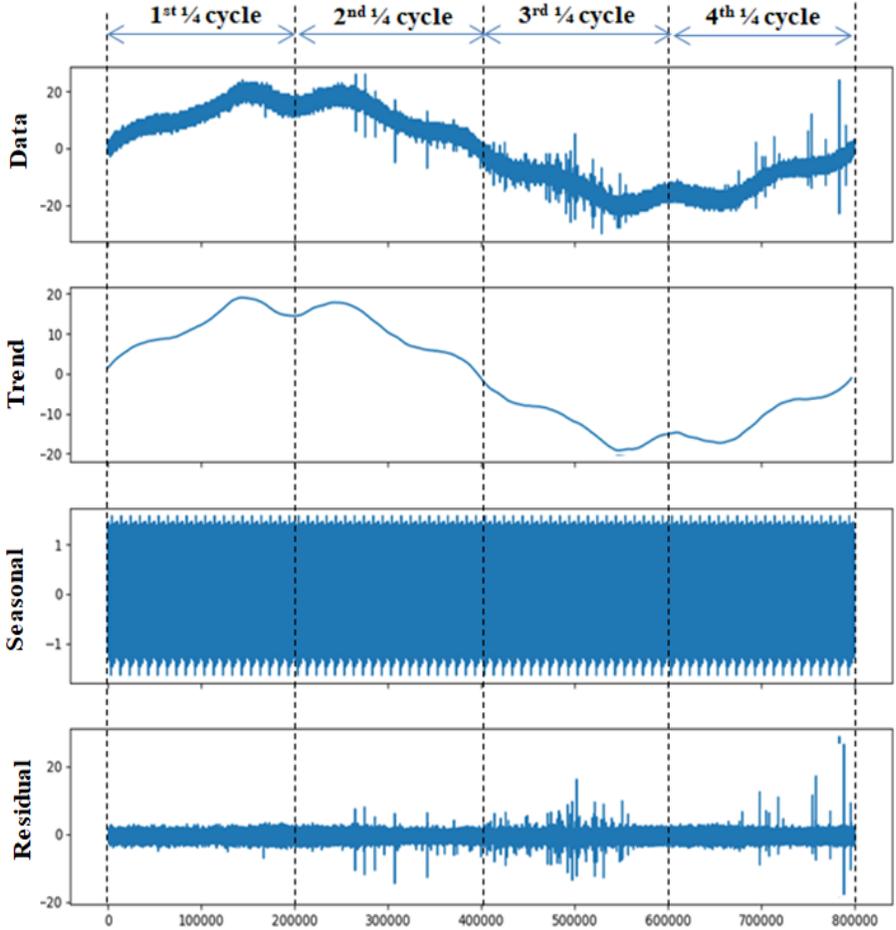

Figure 10. An example of STL decomposition on a signal with PD

*6.2. Results of 4-time Step LSTM Classifier*

Common classification evaluation tools including confusion matrix, precision, recall and F1-Score are used in the evaluation of the LSTM classifier [18]. The confusion matrix of the 4-time step LSTM classifier is shown in Table 1. TP, FP, TN and FN stand for true positive, false positive, true negative and false negative counts.

Table 1: Confusion Matrix for Classification Evaluation

| Total N=1742 | Predicted: Positive(PD signals) | Predicted: Negative(non-PD signals) |
|---|---|---|
| Actual: Positive(PD signals) | TP=758 | FN=177 |
| Actual: Negative(non-PD signals) | FP=193 | TN=614 |

Precision is the ratio of correctly predicted observations to the total predicted observations. Recall is the ratio of correctly predicted observations to the total actual PD or non-PD observations.

For PD signals, the precision index and the recall index are:

$$\begin{cases} Precision = \frac{TP}{TP+FP} \\ Recall = \frac{TP}{TP+FN} \end{cases} \quad (15)$$

For non-PD signals, the precision index and the recall index are:

$$\begin{cases} Precision = \frac{TN}{TN+FN} \\ Recall = \frac{TN}{TN+FP} \end{cases} \quad (16)$$

The F1-Score is the harmonic mean of Precision and Recall:

$$F1 = 2 \frac{Presision \cdot Recall}{Presision + Recall} \quad (17)$$

Table 2 summarizes the precision, recall and F1-Score results following (15)-(17).

Table 2: Precision, Recall and F1-Score for Prediction Evaluation

| Evaluation Category | Precision | Recall | F1-Score |
|---|---|---|---|
| Non-PD Signals | 0.78 | 0.76 | 0.77 |
| PD Signals | 0.80 | 0.81 | 0.80 |
| Average | 0.79 | 0.79 | 0.79 |

*6.3. Comparison of Using Different Numbers of Time Steps*

Figure 11 shows the F1-Scores of the three different numbers of time steps. As can be seen, the 4-time step LSTM classifier provides the best classification results by F1-Score. It is therefore selected as the final classifier for future application. In comparison, 2-time step LSTM is not able to capture enough granular feature variation; 8-time step LSTM may overly amplify less meaningful feature variation and cause overfit which reduces the classification performance.

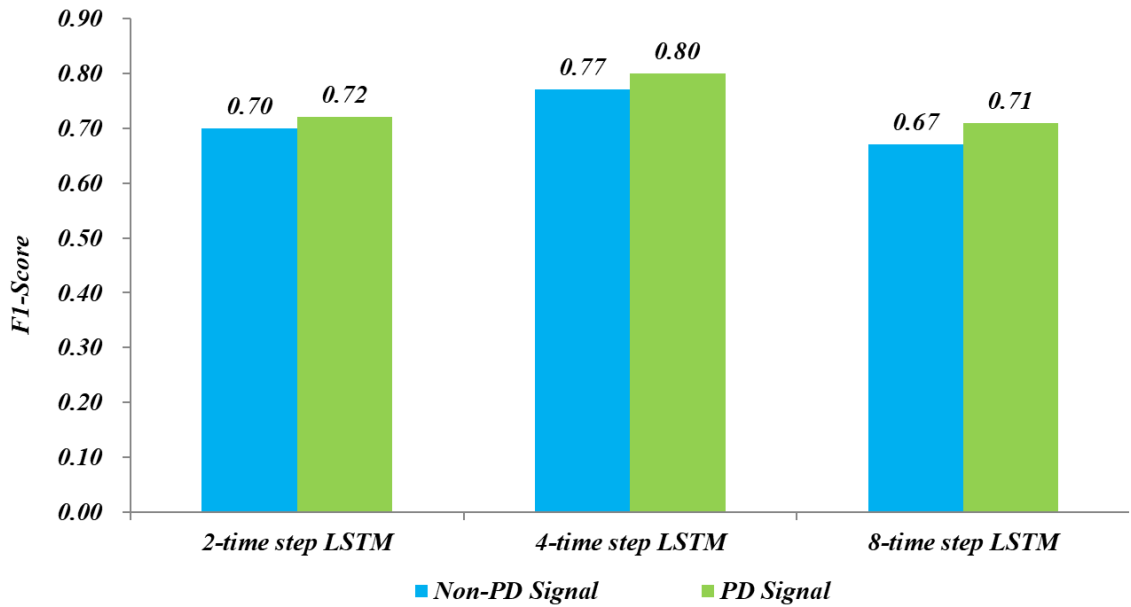
Figure 11. F1-Scores of different time steps

## 6.4. Effects of Noise Reduction and Oversampling

In order to evaluate the effects of noise reduction and oversampling, the F1-Scores of classifiers that only use noise reduction, only use oversampling and use neither of the two techinques are tested and compared in Figure 12. As can be seen, noise reduction and oversampling can both improve the performance of the classification. Oversampling is especially important - without using it, the classification will significantly bias towards non-PD signals and lose its ability to effectively recognize PD signals.

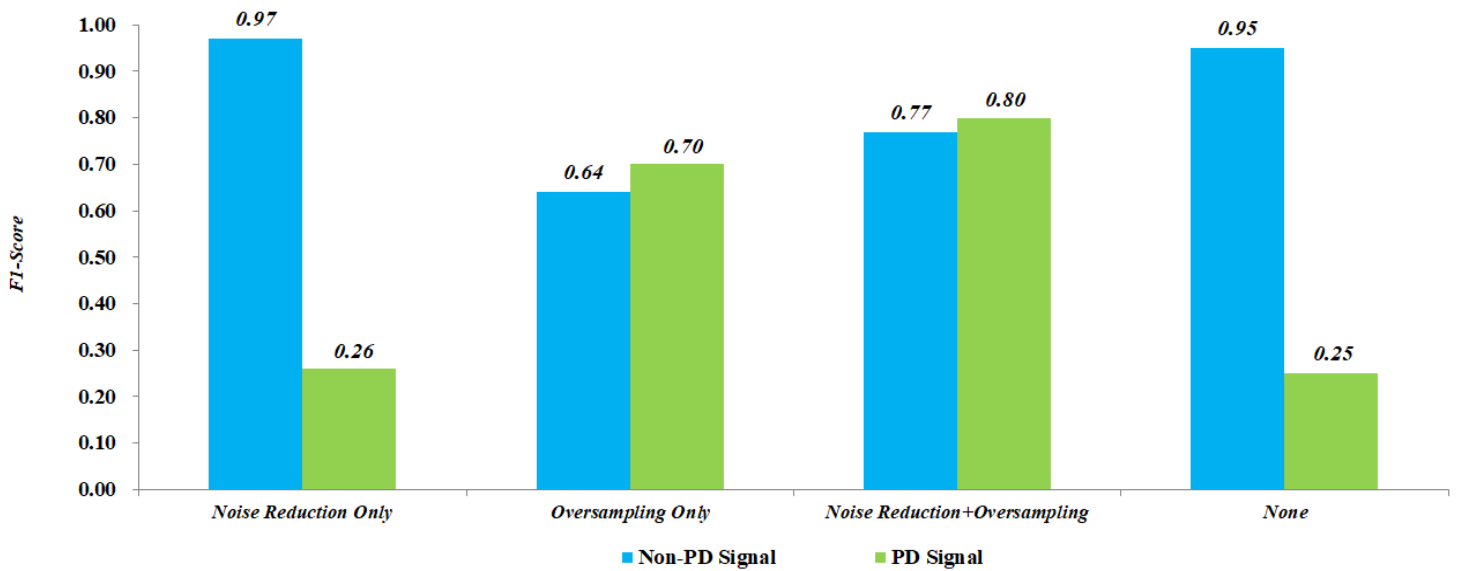
Figure 12. Effect of noise reduction and oversampling

*6.5. Comparison with Other Classifiers*

As part of the model evaluation, the proposed model was compared to various other classifiers established as below. In these methods, the same time-series decomposition method and feature engineering method are applied.

• FNN: A traditional FNN classifier is used to incorporate all the features of a full-cycle. The input layer has 12 neurons and two hidden layers each have 6 neurons. ReLU activation functions are used in the hidden and output layers.

• SVM: SVM classifiers with different kernels are evaluated [19]. The kernels are linear, polynomial (degree 6), Gaussian radial basis function and sigmoid function. It is found out that the SVM with Gaussian radial basis function kernel provided the best classification results by F1-Score in all categories.

• XGBoost [20]: 100 trees are built and the maximum depth of a tree is set to 5.

• Multivariate Logistic Regression (MLR) [21] : A MLR with 12 input variables is used.

As shown in Figure 13, the 4-time step LSTM classifier outperformed all other classifiers.

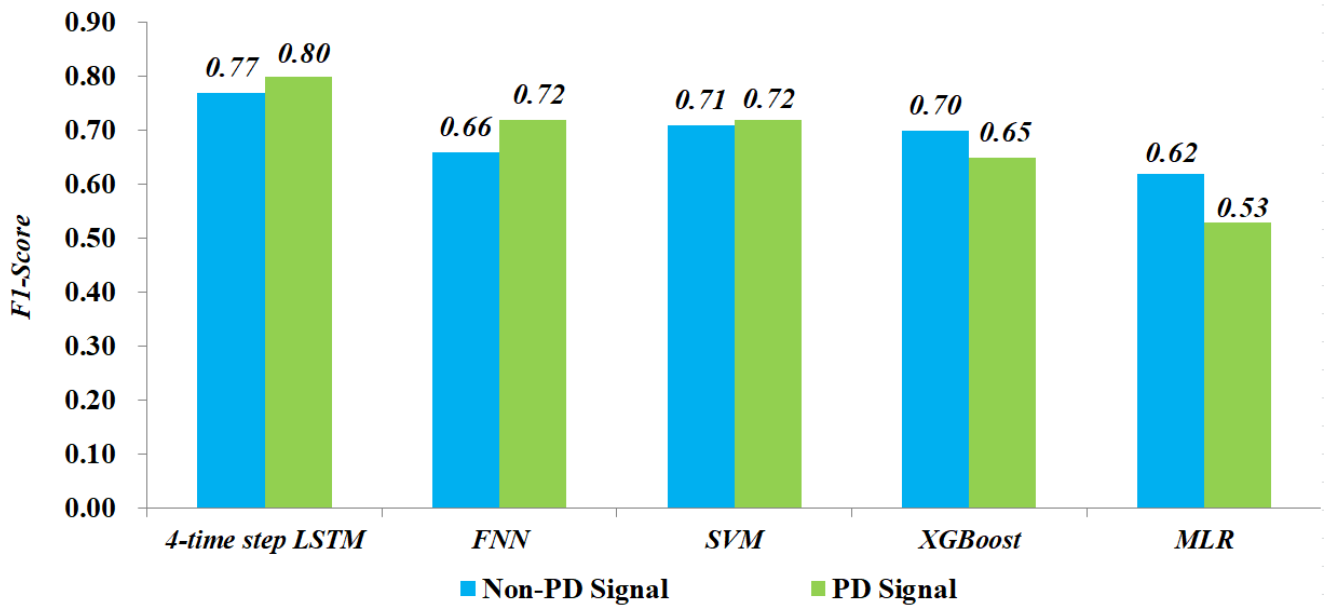

Figure 13. F1-Scores of different classifiers

# 7. Conclusions

This paper developed a novel and comprehensive pattern recognition method based on time-series decomposition and LSTM network in conjunction with a unique feature engineering process to effectively recognize PD activities on aerial covered conductors. The main features of the proposed method are:

- Time-series decomposition can decompose the raw signal and extract the part that is more relevant to partial discharge activities. Four time-series modules with different seasonal–windows lengths are used together to produce the sequential feature vector;

- LSTM network can capture and analyze the sequential characteristics between different time steps within one cycle of signal. This contribution can enhance the PD recognition performance;

- A unique feature engineering process is applied to deal with signal noises and data imbalance problem.

The proposed method was tested on the ENET pubic dataset. Different case studies were conducted and it is found that the 4-time step LSTM classifier with the unique time-series decomposition and feature engineering techniques provided the best classification results measured by F1-Score. The proposed method demonstrated superior performance over various traditional classification methods for PD detection on aerial covered conductors.